\documentclass[twocolumn,amsmath,amsfonts,prl,showpacs]{revtex4}
\usepackage{graphicx}
\usepackage[colorlinks=true,linkcolor=blue,citecolor=blue,urlcolor=blue]{hyperref}

\newcommand{\hP}{{\hat{\Psi}}}
\newcommand{\hPd}{{\hat{\Psi}^{\dagger}}}
\newcommand{\comm}[2]{\left[#1,#2\right]}
\newcommand{\bx}{{\mathbf{x}}}
\newcommand{\hH}{{\hat{H}}}
\newcommand{\dx}{d^{3}x}

\newcommand{\hp}{{\hat{\phi}}}
\newcommand{\hpd}{{\hat{\phi}^{\dagger}}}
\newcommand{\Pn}{{\Psi_{0}}}
\newcommand{\rn}{{\rho_{0}}}
\newcommand{\hbm}{{\frac{\hbar^{2}}{2m}}}

\newcommand{\ii}{{i}}

\newcommand{\pdt}{{\partial_{t}}}
\newcommand{\dk}{{d^{3}k}}
\newcommand{\uk}{{u_{\mathbf{k}}}}
\newcommand{\vk}{{v_{\mathbf{k}}}}
\newcommand{\vks}{{v_{\mathbf{k}}^{*}}}
\newcommand{\ee}{{e}}
\newcommand{\bk}{{\mathbf{k}}}
\newcommand{\ak}{{\hat{a}}_{\bk}}
\newcommand{\akd}{{\hat{a}}^{\dagger}_{\bk}}

\newcommand{\vev}[1]{\langle #1 \rangle_\Omega}
\newcommand{\sev}[1]{\langle #1\rangle_\zeta}

\newcommand{\ie}{{i.e.}}
\newcommand{\eg}{{e.g.}}
\newcommand{\real}{{\mathrm{Re}}}

\begin{document}

\title{Cosmological Constant: A Lesson from Bose-Einstein Condensates}

\author{Stefano Finazzi%
\footnote{Present address: INO-CNR BEC Center and Dipartimento di Fisica, Universit\`a di Trento, via Sommarive 14, 38123
Povo-Trento, Italy.}}
\author{Stefano Liberati}
\affiliation{SISSA, via Bonomea 265, 34136 Trieste, Italy and INFN, sezione di Trieste}
\author{Lorenzo Sindoni}
\affiliation{Albert Einstein Institute, Am M\"uhlenberg 2, 14476 Golm, Germany}

% 04.90.+e 	Other topics in general relativity and gravitation (restricted to new topics in section 04)
% 03.75.Kk 	Dynamic properties of condensates; collective and hydrodynamic excitations, superfluid flow 
% 04.60.Bc 	Phenomenology of quantum gravity
% 95.36.+x 	Dark energy

\begin{abstract}
The cosmological constant is one of the most pressing problems in modern physics.
We address this issue from an emergent gravity standpoint, by using an analogue gravity model.
Indeed, the dynamics of the emergent metric in a Bose-Einstein condensate can be described by a Poisson-like equation with a vacuum source term reminiscent of a cosmological constant.
The direct computation of this term shows that in emergent gravity scenarios this constant may be naturally much smaller than the naive ground-state energy of the emergent effective field theory.
This suggests that a proper computation of the cosmological constant would require a detailed understanding about how Einstein equations emerge from the full microscopic quantum theory.
In this light, the cosmological constant appears as a decisive test bench for any quantum or emergent gravity scenario.
\end{abstract}

\maketitle

\label{sec:intro}

The cosmological constant~\cite{carroll} has been one of the most mysterious and fascinating objects for both cosmologist and theoretical physicists since its introduction almost a century ago~\cite{firstcosm}. Once called by Einstein his greatest blunder, it seems nowadays the driving force behind the current accelerated expansion of the universe.  The explanation of its origin is considered one of the most fundamental issues for our comprehension of general relativity (GR) and quantum field theory.

Since this constant appears in Einstein equations as a source term present even in the absence of matter and with the symmetries of the vacuum ($T_{\mu\nu}^\Lambda\propto g_{\mu\nu}$), it is usually interpreted as a ``vacuum energy''. Unfortunately, this has originated the so-called ``worst prediction'' of physics. In fact, the estimated value, which is naively obtained by integrating the zero-point energies of modes of quantum fields below Planck energy, is about 120 orders of magnitude larger than the measured value. 
Despite the large number of attempts (most notably supersymmetry~\cite{susy,zumino}, which, however, must be broken at low energy) this problem is still open. {We can summarize the situation by saying that, given the absence of custodial symmetries protecting the cosmological term from large renormalization effects, the only option we have to explain observations is fine tuning~\cite{finetuning, rovelli}.}

This huge discrepancy is plausibly due to the use of effective field theory (EFT) calculations for a quantity which can be computed only within a full quantum theory of gravity (see, however, \cite{Paddy} for a proposal in the semiclassical gravity limit). Unfortunately, to date, we do not have any conclusive theory at our disposal. However, the possibility of a failure of our EFT-based intuition is supported by what can be learned from analogue models of gravity~\cite{livrev}, given that, in these models, the way in which the structure of the spacetime emerges from the microscopic theory is fully under control. In~\cite{volovik,volovikbook} it was shown that a naive computation of the ground-state energy using the EFT (the analogue that one would do to compute the cosmological constant), would produce a wrong result. The unique way to compute the correct value seems to use the full microscopic theory. 

Given the deep difference in the structure of the equations of fluid dynamics and GR (and other gravitational theories), an accurate analogy cannot be performed at the dynamical level: indeed, this is forbidden by the absence of diffeomorphism 
invariance and of local Lorentz invariance. However, in~\cite{gls} it was shown for the first time that the evolution of part of the acoustic metric in a Bose-Einstein condensate (BEC) is described by a Poisson equation for a nonrelativistic gravitational field, thus realizing a (partial) dynamical analogy with Newtonian gravity. Noticeably, this equation is endowed with a source term that is naturally identified as a cosmological constant, being there even in the absence of real phonons.

In this Letter we will consider such analogue model for gravity and directly show that the cosmological constant term cannot be computed through the standard EFT approach, confirming the conjecture of~\cite{volovik}. However, we find that also the total ground-state energy of the condensate does not give the correct result: indeed, the cosmological constant is comparable with that fraction of the ground-state energy corresponding to the quantum depletion of the condensate, \ie, to the fraction of atoms inevitably occupying excited states of the single particle Hamiltonian.
In conclusion, the origin and value of such term teach us some interesting lessons about the cosmological constant in emergent gravity scenarios.

%-----------------%
{\it Settings.---}%
%-----------------%
The model used in~\cite{gls} is a modified BEC, including a soft breaking of the $U(1)$ symmetry associated with the conservation of particle number. This unusual choice is a simple trick to give mass to quasiparticles that are otherwise massless by Goldstone's theorem. In second quantization, such a system is described by a canonical field $\hPd$, satisfying
$[\hP(t,\bx),\hPd(t,\bx')]=\delta^3(\bx-\bx'),$
whose dynamics is generated by the grand-canonical Hamiltonian ${\cal\hH}=\hH-\mu\hat N$, where
\begin{multline}
 \hH = \int\! \dx \left[\frac{\hbar^2}{2m} {\nabla \hPd \, \nabla\hP} + V  \hPd\hP 
 \right.\\
 \qquad\qquad\left. + \frac{g}{2}\hPd\hPd\hP\hP-\frac{\lambda}{2}\left(\hP\hP+\hPd\hPd\right)\right],\label{eq:Hsc}
\end{multline}
and $\hat{N}$ is the standard number operator for $\hP$.
In order for the interaction between bosons to be described by $\hH$, the gas must be dilute, i.e., $\rho a^3\ll1$, where $\rho$ is the density and $a\equiv 4\pi g m/\hbar^2$ is the $s$-wave scattering length.
For more details on this model and on possible physical realizations, see~\cite{gls,13_in_gls}. See also~\cite{nbec} for a generalization
to condensates with many components.

We describe the formation of a BEC at low temperature through a
complex function $\Pn$ for the condensate and an operator $\hp$ for the perturbations on top of it~\cite{mp}:
\begin{equation}\label{eq:defphi}
 \hP=\Pn(\mathbb{I}+\hp).
\end{equation}
Clearly, this is only an approximate characterization of the many body ground state. The validity of the mean field approximation must be checked, {\it a posteriori}, by controlling that the fluctuation $\langle \hp^{2} \rangle$ is much smaller than $|\Pn|^2=\rn$. If this is not so, the description of the effective dynamics (\eg\/ the existence of an acoustic geometry where phonons propagate) does not hold any more.
The canonical commutation relation for $\hPd$ implies
\begin{equation}\label{eq:commphi}
 \comm{\hp(t,\bx)}{\hpd(t,\bx')}=\frac{1}{\rn(\bx)}\delta^3(\bx-\bx').
\end{equation}
We adopt the notation of~\cite{fp}, where a rigorous quantization and mode analysis of the field $\hp$ is presented for a standard BEC. Those results are here summarized and generalized to the $U(1)$-breaking case of~\cite{gls}.

For a stationary condensate, $\pdt\Pn=0$ and Eqs.~\eqref{eq:Hsc} and~\eqref{eq:defphi}
lead to a modified Gross-Pita\"evski equation
\begin{equation}
  \left[-\hbm{\nabla}^2+V-\mu+g\rn-\lambda\frac{\Pn^*}{\Pn}\right]\Pn=0. \label{eq:GP}
\end{equation}
For the aim of this Letter, it is enough to consider only
homogeneous backgrounds. Thus, one can assume that $V=0$ and the condensate is at rest,
such that
$\Pn$ has a constant phase. For stability reasons, $\Pn$ must be real ($\Pn^*=\Pn=\sqrt{\rn}$), and Eq.~\eqref{eq:GP} simplifies to
$\mu=g\rn-\lambda$.

The equation for the quasiparticles is solved via Bogoliubov transformation involving the Fourier expansion
\begin{equation}\label{eq:phiexpansion}
\hp
=\int\!\frac{\dk}{\sqrt{\rn(2\pi)^3}}\left[\uk\ee^{-\ii\omega t+\ii{\bf k}\cdot \bx}\ak+\vks\ee^{+\ii\omega t-\ii{\bf k}\cdot \bx}\akd\right],
\end{equation}
where $\ak$ and $\akd$ are quasiparticles' operators and the factor $\sqrt{\rn(2\pi)^3}$ has been inserted such that the Bogoliubov coefficients $\uk$ and $\vk$ obey the standard normalization $|\uk|^2-|\vk|^2=1$.
The dispersion relation is
\begin{equation}\label{eq:disp}
 \hbar^2\omega^2=4\lambda g\rn+\frac{g\rn+\lambda}{m}\hbar^2k^2+\frac{\hbar^4k^4}{4m^2},
\end{equation}
describing massive phonons with ultraviolet corrections, mass $\cal M$, and speed of sound $c_s$~\cite{gls}
\begin{equation}
 {\cal M}=\frac{2\sqrt{\lambda g\rn}}{g\rn+\lambda}m,\qquad c_s^2=\frac{g\rn+\lambda}{m}.\label{eq:csM}
\end{equation}

As shown in~\cite{gls}, when the wavelength is larger than the healing length $\xi=\hbar/mc_s$, phonons propagate in an acoustic geometry with effective local Lorentz invariance and their dispersion relation~\eqref{eq:disp} is relativistic (quadratic). When $k>\xi^{-1}$, the quartic term of the dispersion relation~\eqref{eq:disp} is instead dominant and the effective geometry is not defined. The Lorentz breaking scale $L_{\rm LV}$ is thus identified with $\xi$.

Standard manipulations give $\uk^2=(1-D_\bk^2)^{-1}$ and $\vk^2=D_\bk^2\uk$, where $\uk$ and $\vk$ are chosen to be real and
\begin{equation}
 D_\bk\equiv\frac{\hbar\omega-\left(\hbar^2 k^2/2m+g\rn+\lambda\right)}{g\rn-\lambda}.\label{eq:dk}
\end{equation}
%

\begin{figure}[b]
 \includegraphics[width=0.87\columnwidth]{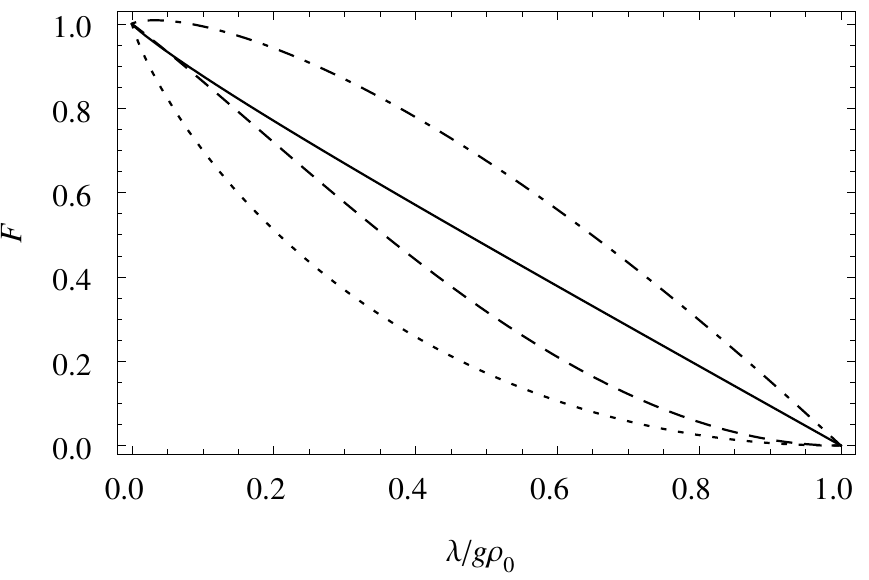}
 \caption{\label{fig:fs}$F_h$ [dashed line, Eq.~\eqref{eq:h2vev2}], $F_\rho$ [dotted line, Eq.~\eqref{eq:depletion}], $F_{\phi\phi}$ [dot-dashed line, Eq.~\eqref{eq:phiphi}], and $F_{\Lambda}$ [solid line, Eq.~\eqref{eq:lambda}].}
\end{figure}

%----------------------------------%
{\it Vacuum expectation values.---}%
%----------------------------------%
We can now compute the vacuum expectation value of $\cal\hH$ in the ground state $|\Omega \rangle$, the Fock vacuum of the quasiparticles ($\ak|\Omega\rangle=0, \: \forall\, \bk$). To this aim, it is convenient to expand $\cal\hH$ in powers of $\hp$: ${\cal\hH}\approx{\cal H}_0+{\cal\hH}_1+{\cal\hH}_2$, where ${\cal H}_0$, ${\cal\hH}_1$, and ${\cal\hH}_2$ contain, respectively, no power of $\hp$, only first powers, and only second powers, and higher order terms associated with quasiparticles' self-interactions are neglected.
The energy density $h_0$ of the condensate (density of ${\cal H}_0$) and the density $h_2$ of the expectation value of ${\cal\hH}_2$ are
\begin{equation}\label{eq:h02vev}
 h_0=-\frac{g\rn^2}{2},\qquad h_2=-\int\!\frac{\dk}{(2\pi)^3}\hbar\omega|\vk|^2,
\end{equation}
while the expectation value of ${\cal\hH}_1$ vanishes because it contains only odd powers of $\ak$ and $\akd$.
The integral in Eq.~\eqref{eq:h02vev} is computed by using the above given expression
for $\vk$. Applying standard regularization techniques~\cite{someonecm}
\begin{equation}\label{eq:h2vev2}
 h_2=\frac{64}{15\sqrt{\pi}}g\rn^2\sqrt{\rn a^3}\,\,F_h\!\left(\frac{\lambda}{g\rn}\right),
\end{equation}
where $F_h$ is plotted in Fig.~\ref{fig:fs} (dashed line) and $F_h(0)$=1.

The total grand-canonical energy density is therefore
\begin{equation}\label{eq:grandpotential}
 h=h_0+h_2=\frac{g\rn^2}{2}\left[-1+\frac{128}{15\sqrt{\pi}}\sqrt{\rn a^3}\,\,F_h\!\left(\frac{\lambda}{g\rn}\right)\right]
\end{equation}
and it coincides with the well known Lee-Huang-Yang formula~\cite{lhy} when $\lambda=0$.
 
The number density operator $\hat N$ is analogously expanded in powers of $\hp$: $\hat N=N_0+\hat N_1+\hat N_2$.
The density of $N_0$ is $\rho_0=|\Pn|^2$, $\vev{\hat N_1}=0$, and $\rho_2\equiv\vev{\hat N_2}$ is
\begin{equation}\label{eq:depletion}
 \rho_2=\rn\vev{\hpd\hp}=\!\int\!\!\frac{\dk}{(2\pi)^3}|\vk|^2=\frac{8\rn}{3\sqrt{\pi}}\sqrt{\rn a^3}\,\,F_\rho\!\left(\frac{\lambda}{g\rn}\right),
\end{equation}
where $F_\rho$ satisfies $F_\rho(0)=1$ (see Fig.~\ref{fig:fs}, dotted line).
This is the number density of noncondensed atoms ({\it depletion}) and it is basically the magnitude of the fluctuations around the mean field.
Note that $\rn a^3\ll 1$, as described after Eq.~\eqref{eq:Hsc}.

Furthermore, when $\lambda=0$, inverting the expression for total particle density, $\rho=\rn+\rho_2$, one obtains, up to the first order in $\sqrt{\rho a^3}$
\begin{equation}
 \rn=\rho\left[1-\frac{8}{3\sqrt{\pi}}\sqrt{\rho a^3}\right],
\end{equation}
which is the density of condensed atoms in terms of the total density~$\rho$ and the scattering length~$a$~\cite{lhy}. In this case, $\mu=g\rn$, such that the energy density $\epsilon$ (density of $\vev{\hH}=\vev{{\cal H}+\mu\hat N}$) is
\begin{equation}
 \epsilon=h+\mu\rho= \frac{g\rho^2}{2}\left[1+\frac{128}{15\sqrt{\pi}}\sqrt{\rho a^3}\right].\label{eq:energy}
\end{equation}
This is the well known Lee-Huang-Yang~\cite{lhy} formula for the ground-state energy in a condensate at zero temperature.
In general, when the $U(1)$ breaking term is small, this term is expected to be the dominant contribution to the ground-state energy of the condensate.

%---------------------------------------%
{\it Analogue cosmological constant.---}%
%---------------------------------------%
When the homogeneous condensate background is perturbed by small inhomogeneities, the Hamiltonian for the quasiparticles can be written as (see~\cite{gls})
\begin{equation}\label{eq:nonrelhquasip}
\hH_{\rm quasip.} \approx {\cal M} c_s^2- \frac{\hbar^2 \nabla^2}{2{\cal M}} + {\cal M}\Phi_{g}.
\end{equation}
$\hH_{\rm quasip.}$ is the nonrelativistic Hamiltonian for particles of mass ${\cal M}$ [see Eq.~\eqref{eq:csM}] in a gravitational potential
\begin{equation}\label{gravitationalpotential}
\Phi_{g}(\bx) = \frac{(g\rn+3\lambda)(g\rn+\lambda)}{2\lambda m} u(\bx)
\end{equation}
and $u(\bx)=[(\rn(\bx)/\rho_\infty)-1]/2$, where $\rho_\infty$ is the asymptotic density of the condensate.
Moreover, the dynamics of the potential $\Phi_{g}$ is described by a Poisson-like equation
\begin{equation}
 \left[\nabla^2-\frac{1}{L^2}\right]\Phi_{g}=4\pi G_{\rm N}\rho_{p}+C_\Lambda,\label{eq:poisson}
\end{equation}
which is the equation for a nonrelativistic short-range field with length scale $L$ and gravitational constant $G_{\rm N}$:
\begin{equation}\label{eq:GN}
 L=\frac{a}{\sqrt{16\pi\rn a^3}},\quad
 G_{\rm N}=\frac{g(g\rn+3\lambda)(g\rn+\lambda)^2}{4\pi\hbar^2m\lambda^{3/2}(g\rn)^{1/2}}.
 \end{equation}
Despite the obvious difference between $\Phi_{g}$
and the usual Newtonian gravitational potential, we insist in calling it the {\it Newtonian potential}
because it enters the acoustic metric exactly as
the Newtonian potential enters the metric tensor  in the Newtonian limit of GR.
The appearance of a short-range interaction in Eq.(\ref{eq:poisson}) is an artifact of the model.
In \cite{nbec} it has been shown how to obtain a long range analogue gravitational potential in a spinor BEC. However, the reasoning is identical in all the other relevant aspects, and the key result is unchanged.

The source term in Eq.~\eqref{eq:poisson} contains both the contribution of real phonons (playing the role of matter)
\begin{equation}
 \rho_{p}  = {\cal M}\rn\left[\left(\sev{\hpd\hp}-\vev{\hpd\hp}\right)+
 \frac{1}{2}\real\left(\sev{\hp\hp}-\vev{\hp\hp}\right)\right],\label{eq:rhophonons}
\end{equation}
where $|\zeta\rangle$ is some state of real phonons, as well as
a cosmological constant like term (present even in the absence of phonons/matter)
\begin{equation}\label{eq:Clambda}
 C_\Lambda=\frac{2g\rn (g\rn+3\lambda)(g\rn+\lambda)}{\hbar^2\lambda}
 \real \left[\vev{\hpd\hp}+\frac{1}{2}\vev{\hp\hp}\right].
\end{equation}
Note that the source term in the correct weak field approximation of Einstein equations is $4\pi G_{\rm N}(\rho+3p/c^2)$. For standard nonrelativistic matter, $p/c^2$ is usually negligible with respect to $\rho$. However, it cannot be neglected for the cosmological constant, since $p_\Lambda/c^2=-\rho_\Lambda$. As a consequence  $C_\Lambda=-2c_{s}^{2}\Lambda$, where $\Lambda$ would be the GR cosmological constant.
From Eq.~\eqref{eq:depletion} and evaluating
\begin{equation}\label{eq:phiphi}
 \vev{\hp\hp}=\!\int\!\!\frac{\dk}{\rn(2\pi)^3}\uk\vk=\frac{8}{\sqrt{\pi}}\sqrt{\rn a^3}\,F_{\phi\phi}\!\left(\frac{\lambda}{g\rn}\right),
\end{equation}
where $F_{\phi\phi}(0)=1$ (see Fig.~\ref{fig:fs}, dot-dashed line), we obtain
\begin{equation}\label{eq:lambda}
 \Lambda=-\frac{20m\,g\rn\,(g\rn+3\lambda)}{3\sqrt{\pi}\hbar^2\lambda}\sqrt{\rn a^3}\,F_\Lambda\!\left(\frac{\lambda}{g\rn}\right),
\end{equation}
where $F_\Lambda=(2F_\rho+3F_{\phi\phi})/5$ (see Fig.~\ref{fig:fs}, solid line).

Let us now compare the value of $\Lambda$ either with the ground-state grand-canonical energy density $h$ [Eq.~\eqref{eq:grandpotential}], which in~\cite{volovik} was suggested as the correct vacuum energy corresponding to the cosmological constant, or with the ground-state energy density $\epsilon$ of Eq.~\eqref{eq:energy}.
Evidently, $\Lambda$ does not correspond to either of them: even when taking into account the correct behavior at small scales, the vacuum energy computed with the phonon EFT does not lead to the correct value of the cosmological constant appearing in Eq.~\eqref{eq:poisson}.
Noticeably, since $\Lambda$ is proportional to $\sqrt{\rn a^3}$, it can even be arbitrarily smaller both than $h$ and than $\epsilon$, if the condensate is very dilute. Furthermore, $\Lambda$ is proportional only to the subdominant second order correction of $h$ or $\epsilon$, which is strictly related to the depletion [see Eq.~\eqref{eq:depletion}].

%---------------------------%
{\it Fundamental scales.---}%
%---------------------------%
Several scales show up in this system, in addition to the naive Planck scale computed by combining $\hbar$ and the emergent constants $G_{\rm N}$ and $c_s$:
\begin{equation}
 L_{\rm P}=\sqrt{\frac{\hbar c_s^5}{G_{\rm N}}}\propto \left(\frac{\lambda}{g\rn}\right)^{-3/4}(\rn a^3)^{-1/4} a.
\end{equation}
For instance, the Lorentz-violation scale $L_{\rm LV}=\xi \propto(\rn a^3)^{-1/2} a$ differs from $L_{\rm P}$, suggesting that the breaking of the Lorentz symmetry might be expected at scale much longer than the Planck length (energy much smaller than the Planck energy), since the ratio $L_{\rm LV}/L_{\rm P}\propto(\rn a^3)^{-1/4}$ increases with the diluteness of the condensate. 

Note that $L_{\rm LV}$ scales with $\rn a^3$ exactly as the range of the gravitational force [see Eq.~\eqref{eq:GN}], signaling that this model is too simple to correctly grasp all the desired features. However,
in more complicated systems~\cite{nbec}, this pathology can be cured, in the presence of suitable symmetries, leading to long range potentials.

It is instructive to compare the energy density corresponding to $\Lambda$ to the Planck energy density:
\begin{equation}
{\cal E}_\Lambda=\frac{\Lambda c_s^4}{4\pi G_{\rm N}},\quad
{\cal E}_{\rm P}=\frac{c_s^7}{\hbar G_{\rm N}^2},\quad
 \frac{{\cal E}_\Lambda}{{\cal E}_{\rm P}}\propto \rn a^3\left(\frac{\lambda}{g\rn}\right)^{-5/2}.
\end{equation}
The energy density associated with the analogue cosmological constant is 
much smaller than the values computed from zero-point-energy calculations with a cutoff at the Planck scale. Indeed, the ratio between these two quantities is controlled by the diluteness parameter $\rn a^3$.

%----------------------%
{\it Final remarks.---}%
%----------------------%
Taken at face value, this relatively simple model displays too many crucial differences with any realistic theory of gravity to provide conclusive evidences. However, it displays an alternative path to the cosmological constant,
from the perspective of a microscopic model. The analogue cosmological constant that we have discussed {\it cannot} be computed as the total zero-point energy of the condensed matter system, even when taking into account the natural cutoff coming from the knowledge of the microphysics~\cite{volovik}. In fact the value of $\Lambda$ is related only to the (subleading) part of the zero-point energy proportional to the quantum depletion of the condensate. This holds also in a spinor BEC model, since the reasoning there is absolutely identical.
The virtue of the single BEC model is to display the key physical result without obscuring it with unnecessary mathematical complications, without loss of generality.
Interestingly, this result finds some support from arguments within loop quantum gravity models \cite{Alexander}, suggesting a BCS energy gap as a (conceptually rather different) origin for the cosmological constant.

The implications for gravity are twofold. First, there could be no {\it a priori} reason why the cosmological constant should be computed as the zero-point energy of the system. More properly, its computation must inevitably pass through the derivation of Einstein equations emerging from the underlying microscopic system.
Second, the energy scale of $\Lambda$ can be several orders of magnitude smaller than all the other energy scales for the presence of a very small number, nonperturbative in origin, which cannot be computed within the framework of an EFT
dealing only with the emergent degrees of freedom (\ie, semiclassical gravity).

The model discussed in this Letter shows all this explicitly: the energy scale of $\Lambda$ is here lowered by the diluteness parameter of the condensate.
Furthermore, our analysis strongly supports a picture where gravity is a collective phenomenon in a pregeometric theory.
In fact, the cosmological constant puzzle is elegantly solved in those scenarios.
From an emergent gravity approach, the low energy effective action (and its renormalization group flow) is computed within a framework that has nothing to do with quantum field theories in curved spacetime.
Indeed, if we interpreted the cosmological constant as a coupling constant controlling some self-interaction of the gravitational field, rather than as a vacuum energy, it would immediately follow that the explanation of its value (and of its properties under renormalization) would naturally sit outside the domain of semiclassical gravity.

For instance, in a group field theory scenario (a generalization to higher dimensions of matrix models for two dimensional quantum gravity \cite{GFT}), it is transparent that the origin of the gravitational coupling constants has nothing to do with ideas like ``vacuum energy'' or statements like ``energy gravitates'', because energy {\it itself} is an emergent concept. Rather, the value of $\Lambda$ is determined by the microphysics, and, most importantly, by the procedure to approach the continuum semiclassical limit.
In this respect, it is conceivable that the very notion of cosmological constant as a form of energy intrinsic to the vacuum is ultimately misleading.
To date, little is known about the macroscopic regime of models like group field theories, even though some preliminary steps have been recently done~\cite{Oriti:2010hg}.
Nonetheless, analogue models elucidate in simple ways what is expected to happen and can suggest how to further develop investigations in quantum gravity models.
In this respect, the reasoning of this Letter sheds a totally different light on the cosmological constant problem, turning it from a failure of effective field theory to a question about the emergence of the spacetime.

\begin{acknowledgments}
 The authors wish to thank C. Barcel\'o, P. Jain, G. Jannes, R. Parentani, S. Sonego, R. Sch\"utzhold, W. G. Unruh and S. Weinfurtner for stimulating discussions.
\end{acknowledgments}

\end{document}